\documentclass[%
superscriptaddress,
preprint,
amsmath,amssymb,
aps,
prx,
]{revtex4-1}

\usepackage{array}
\usepackage{graphicx}
\usepackage{dcolumn}
\usepackage{bm}
\usepackage{amsmath}
\usepackage{makecell}

\usepackage{algorithm}
\usepackage{algpseudocode}

\begin{document}

\title{On-the-Fly Active Learning of Interpretable Bayesian Force Fields for Atomistic Rare Events}


\author{Jonathan Vandermause}
\affiliation{Department of Physics, Harvard University, Cambridge, MA 02138, USA}
\affiliation{John A. Paulson School of Engineering and Applied
Sciences, Harvard University, Cambridge, MA 02138, USA}

\author{Steven B. Torrisi}
\affiliation{Department of Physics, Harvard University, Cambridge, MA 02138, USA}

\author{Simon Batzner}
\affiliation{John A. Paulson School of Engineering and Applied
Sciences, Harvard University, Cambridge, MA 02138, USA}
\affiliation{Center for Computational Engineering, Massachusetts Institute of Technology, Cambridge, MA 02139, USA}

\author{Yu Xie}
\affiliation{John A. Paulson School of Engineering and Applied
Sciences, Harvard University, Cambridge, MA 02138, USA}

\author{Lixin Sun}
\affiliation{John A. Paulson School of Engineering and Applied
Sciences, Harvard University, Cambridge, MA 02138, USA}

\author{Alexie M. Kolpak}
\affiliation{Department of Mechanical Engineering, Massachusetts Institute of Technology, Cambridge, MA 02139, USA}

\author{Boris Kozinsky}
\affiliation{John A. Paulson School of Engineering and Applied
Sciences, Harvard University, Cambridge, MA 02138, USA}
\affiliation{Bosch Research, Cambridge, MA 02139, USA}

\date{\today}

\begin{abstract}
    Machine learned force fields typically require manual construction of training sets consisting of thousands of first principles calculations, which can result in low training efficiency and unpredictable errors when applied to structures not represented in the training set of the model. 
    This severely limits the practical application of these models in systems with dynamics governed by important rare events, such as chemical reactions and diffusion. We present an adaptive Bayesian inference method for automating the training of interpretable, low-dimensional, and multi-element interatomic force fields using structures drawn on the fly from molecular dynamics simulations. Within an active learning framework, the internal uncertainty of a Gaussian process regression model is used to decide whether to accept the model prediction or to perform a first principles calculation to augment the training set of the model. The method is applied to a range of single- and multi-element systems and shown to achieve a favorable balance of accuracy and computational efficiency, while requiring a minimal amount of \textit{ab initio} training data.
    We provide a fully open-source implementation of our method, as well as a procedure to map trained models to computationally efficient tabulated force fields. 
\end{abstract}


\maketitle

\section{Introduction}

Recent machine learned (ML) force fields have been shown to achieve high accuracy for a number of molecular and solid-state systems \cite{szlachta2014accuracy, behler2011neural, thompson2015spectral, deringer2017machine, chmiela2017machine, schutt2017schnet, deringer2018data, bartok2018machine, zhang2018deep, zhang2018end, smith2019approaching}. These methods provide a promising path toward long, large-scale molecular dynamics (MD) simulations driven by force predictions that approach the accuracy of quantum mechanical methods like density functional theory (DFT). However, most currently available ML force fields return point estimates of energies, forces, and stresses rather than predictive distributions that reflect model uncertainty, making the incorporation of accurate uncertainty estimates into ML force fields an outstanding challenge \cite{botu2016machine, podryabinkin2017active, mishra2018multiobjective, janet2019quantitative, musil2019fast, zhang2019active, podryabinkin2019accelerating}. Without model uncertainty, a laborious fitting procedure is required, which usually involves manually or randomly selecting thousands of reference structures from a database of first principles calculations. In production MD runs, a lack of principled means to compute predictive uncertainties makes it difficult to determine when the force field is trustworthy, leading to unreliable results and lack of guidance on how to update the model in the presence of new data. 

Here, we show that active learning based on Gaussian process (GP) regression can accelerate and automate the training of high-quality force fields by making use of accurate internal estimates of model error. By combining DFT with low-dimensional GP regression models during molecular dynamics simulations, accurate force fields for a range of single- and multi-element systems are obtained with $\sim 100$ DFT calculations. Moreover, we demonstrate that the model can be flexibly and automatically updated when the system deviates from previous training data. Such a reduction in the computational cost of training and updating force fields promises to extend ML modeling to a much wider class of materials than has been possible to date. The method is shown to successfully model rapid crystal melts and rare diffusive events, and so we call our method FLARE: Fast Learning of Atomistic Rare Events, and make the open-source software freely available online \cite{flare}.

The key contribution of this work that makes on-the-fly learning possible is the development of a fully interpretable low-dimensional and nonparametric force field that provides trustworthy estimates of model uncertainty. Typical ML force fields involve regression over a high-dimensional descriptor space chosen either on physical grounds \cite{behler2011atom, bartok2013representing} or learned directly from \textit{ab initio} data \cite{schutt2017schnet, zhang2018end}. These approaches involve highly flexible models with many physically non-interpretable parameters, complicating the task of inferring a posterior distribution over model parameters. We instead bypass the need for a high dimensional descriptor by imposing a physical prior that constrains the model to $n$-body interactions, with high accuracy observed in practice with 2- and 3-body models. Because the low-dimensional descriptor space of our models can be sampled with a small amount of training data, our method avoids sparsification, a procedure that is used in Gaussian approximation potentials to make inference tractable with many-body descriptors like SOAP \cite{bartok2010gaussian, bartok2013representing, bartok2015gaussian}, but that requires approximate treatment of GP uncertainty estimates \cite{williams2006gaussian,quinonero2005unifying}. The learning task is simplified as a result, making it possible to automatically tune the model's hyperparameters in a data-driven fashion and derive trustworthy estimates of model uncertainty. This opens the door to a practical uncertainty-driven method for selecting training points ``on the fly'' \cite{li2015molecular}, allowing an accurate force field to be trained with a minimal number of relatively expensive first principles calculations.

The resulting GP-based force fields are interpretable in three important respects. First, the underlying energy model of the GP is a physically motivated sum over $n$-body contributions, such that each cluster of $n-1$ neighbors in an atom's environment makes a direct contribution to the force on that atom. This establishes a connection to previous physically motivated force fields, most notably the Stillinger-Weber force field \cite{stillinger1985computer}, which also sums over 2- and 3-body contributions but is limited to a specific analytic form. Our models, by contrast, learn nonparametric 2- and 3-body functions directly from \textit{ab initio} data, allowing the models to generalize well to complex multi-element systems, as we show in the Results section below. Second, the model does not require a descriptor of the entire local environment of an atom, instead relying on a kernel that directly compares interatomic distances of small clusters of atoms. As a result, the only free parameters in the model are a small set of hyperparameters of the GP kernel function, each of which has a direct interpretation and can be rigorously optimized by maximizing the log marginal likelihood of the training data.  Neural network and Gaussian approximation potentials, on the other hand, rely on complex high-dimensional descriptors of an atom's environment, making it less apparent how the force acting on an atom is related to the configuration of its neighbors. Finally, and most importantly for active learning, the uncertainty estimates of our GP models break down into two contributions: the epistemic uncertainty $\sigma_{i\alpha}$, which is assigned to each atom $i$ and force component $\alpha$ and is determined by distance from the training set, and the noise uncertainty $\sigma_n$, which characterizes fundamental variability in the training data that cannot be captured by the model. The latter source of error arises from several simplifying approximations that improve computational efficiency, including the exclusion of interactions outside the cutoff radius of the model, the decomposition of global energies into local contributions, and the restriction to 2- and 3-body interactions \cite{bartok2015gaussian, deringer2017machine}. By optimizing the noise uncertainty $\sigma_n$ of the GP, the combined magnitude of these errors can be learned directly from the data (see Methods). The interpretable uncertainties derived from the GP model provide a principled basis for automated training, in which a local environment is added to the training set of the model when the epistemic uncertainty $\sigma_{i\alpha}$ on a force component exceeds a chosen multiple of the noise uncertainty $\sigma_n$.

Other GP and active learning based methods for force field training have been proposed in the literature, and we discuss them briefly here to place our method in context. Bart\'{o}k \textit{et al.}\ pioneered the use of GP-based force fields in the Gaussian approximation potential (GAP) framework \cite{bartok2010gaussian,bartok2015gaussian}, with subsequent applications combining 2- and 3-body descriptors with the many-body SOAP kernel to achieve high accuracy for a range of extended systems \cite{bartok2013representing, deringer2017machine,deringer2018data}. Recent GAP studies have reported uncertainty estimates on local energy predictions \cite{bartok2018machine} and introduced self-guided protocols for learning force fields based on random structure searching rather than uncertainty-driven active learning \cite{deringer2018data, bernstein2019novo}. Rupp \textit{et al.}\ \cite{rupp2014machine} and more recently Uteva \textit{et al.}\ \cite{uteva2018active} used GP regression to model potential energy surfaces of small molecular systems with active learning, and Smith \textit{et al.}\ recently proposed a query-by-committee procedure for actively learning neural network force fields for small molecules \cite{smith2018less}. On-the-fly force field training for extended systems was first proposed by Li, Kermode, and De Vita \cite{li2015molecular}, but the method relied on performing DFT calculations to evaluate model error due to a lack of correlation between the internal error of their GP model and true model error \cite{li2014fly}. Podryabinkin and Shapeev developed an on-the-fly method for their linear moment tensor potentials \cite{shapeev2016moment} using the D-optimality criterion, which provides an internal information-theoretic measure of distance from the training set \cite{podryabinkin2017active}, with subsequent applications to molecules, alloys, and crystal structure prediction \cite{podryabinkin2019accelerating, gubaev2019accelerating, gubaev2018machine}. The D-optimality criterion is usually restricted to linear models and does not provide direct error estimates on model predictions. More recently, Jinnouchi \textit{et al.}\ combined a multi-element variant of the SOAP kernel with Bayesian linear regression to obtain direct Bayesian error estimates on individual force components, which was used to perform on-the-fly training of force fields to study melting points and perovskite phase transitions \cite{jinnouchi2019fly, jinnouchi2019phase}. This approach relies on a decomposition of the atomic density of each atom into many-body descriptors based on spherical Bessel functions and spherical harmonics, with the number of descriptors growing quadratically with the number of elements in the system \cite{de2016comparing}.
The machine learned force fields presented here possess four important features that have not been simultaneously achieved before: they are nonparametric, fully Bayesian, explicitly multi-element, and can be mapped to highly efficient tabulated force fields, making our automated method for training these models widely applicable to a range of complex materials.


\section{Results} \label{results}

\subsection{FLARE: An on-the-fly learning method} \label{alg}

\begin{figure*}
	\centering
	\includegraphics[width=6.48in]{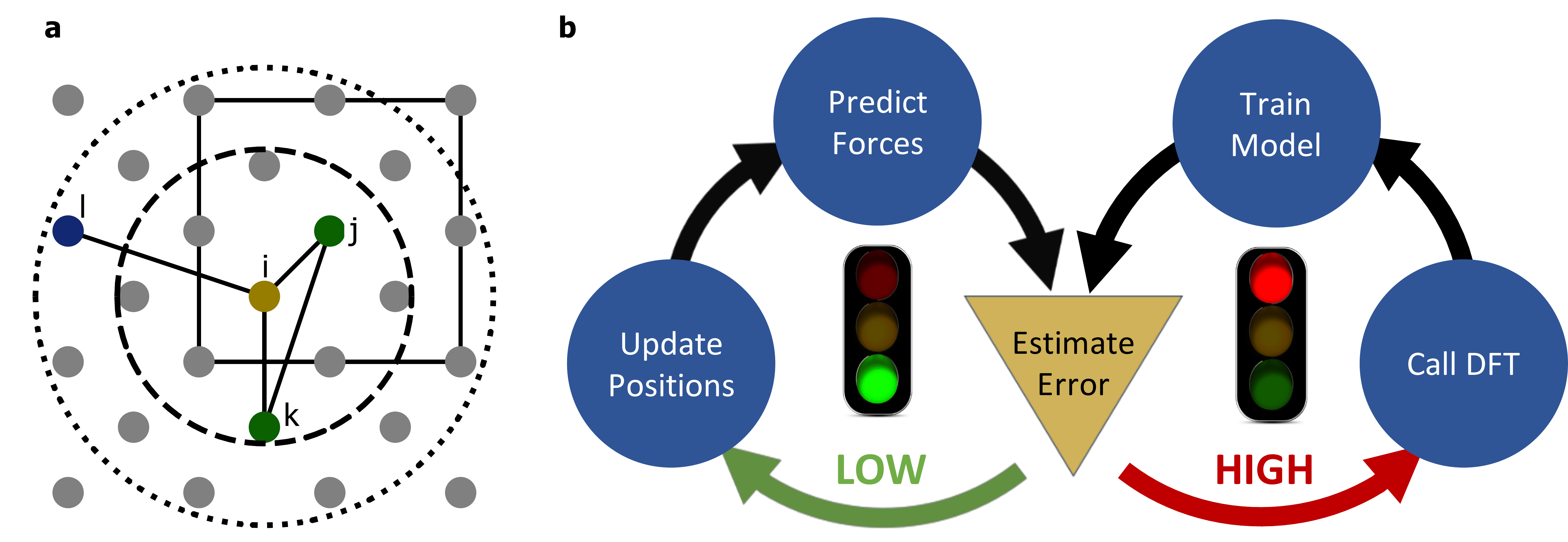}
	\caption{Fast learning of atomistic rare events (FLARE): an on-the-fly learning method for automatically training force fields. \textbf{a} The 2- and 3-body multi-element kernels used in this work. The local environment of the central atom (gold) consists of all atoms within the 2- and 3-body cutoff spheres (dotted and dashed lines, respectively), including images of atoms in the primary periodic cell (solid square). The kernel is calculated by comparing clusters of two and three atoms of the same type, as determined by the chemical species of the atoms in the cluster. \textbf{b} An overview of the on-the-fly learning algorithm. Left loop: Molecular dynamics steps are proposed by the current GP force field, with the epistemic uncertainties $\sigma_{i\alpha}$ on all force components  monitored at each step. Right loop: If the epistemic uncertainty on a force component rises above a chosen multiple of the optimized noise uncertainty $\sigma_n$ of the GP, DFT is called and the training set of the GP is updated with the highest uncertainty local environments.}
	\label{cut}
\end{figure*}

The goal of FLARE is to automate the training of accurate and computationally efficient force fields that can be used for large-scale molecular dynamics simulations of multi-element systems. The low-dimensional GP kernel that we use throughout this work, sketched in Fig.\ \ref{cut}(a), is calculated by comparing interatomic distances of clusters of two and three atoms, similar to the single-element kernel presented in Ref.\ \cite{glielmo2018efficient} but here generalized to arbitrarily many chemical species. If the two clusters are not of the same type, as determined by the chemical species of the atoms in the cluster, the kernel is assigned a value of zero, allowing the GP to differentiate between chemical species while remaining low dimensional (see Methods). Restricting the model to a sum over two- and three-dimensional contributions reduces the cost of training the model, allowing the descriptor space to be systematically sampled with a relatively small number of DFT calculations, and also reduces the cost of production MD runs with the final trained model, since the GP can be mapped onto efficient cubic spline models that allow the 2- and 3-body contributions to the force on an atom to be directly evaluated \cite{glielmo2018efficient}. We have implemented this mapping as a pair style in the molecular dynamics software LAMMPS, allowing us to study multi-element systems containing more than ten thousand atoms over nanosecond timescales (Fig.\ \ref{iod} below).

The low dimensionality of our models also makes it practically feasible to rigorously optimize the hyperparameters of the kernel function, which leads to trustworthy estimates of model uncertainty. The reliability of these uncertainties is the key feature of our approach that enables FLARE, an adaptive method for training force fields on the fly during molecular dynamics. As sketched in Fig.\ \ref{cut}b, the algorithm takes an arbitrary structure as input and begins with a call to DFT, which is used to train an initial GP model on the forces acting on an arbitrarily chosen subset of atoms in the structure. The GP then proposes an MD step by predicting the forces on all atoms, at which point a decision is made about whether to accept the predictions of the GP or to perform a DFT calculation. The decision is based on the epistemic uncertainty $\sigma_{i\alpha}$ of each GP force component prediction (defined in Eq.\ (\ref{reg}) of Methods), which estimates the error of the prediction due to dissimilarity between the atom's environment and the local environments stored in the training set of the GP. In particular, if any $\sigma_{i\alpha}$ exceeds a chosen multiple of the current noise uncertainty $\sigma_n$ of the model, a call to DFT is made and the training set is augmented with the forces acting on the $\mathcal{N}_{\text{added}}$ highest uncertainty local environments, the precise number of which can be tuned to increase training efficiency. All hyperparameters, including the noise uncertainty $\sigma_n$, are optimized whenever a local environment and its force components are added to the training set, allowing the error threshold to adapt to novel environments encountered during the simulation (see Methods). 


\subsection{Characterization of model uncertainty} \label{uncert}
\begin{figure*}
	\centering
    \includegraphics[width=6.48in]{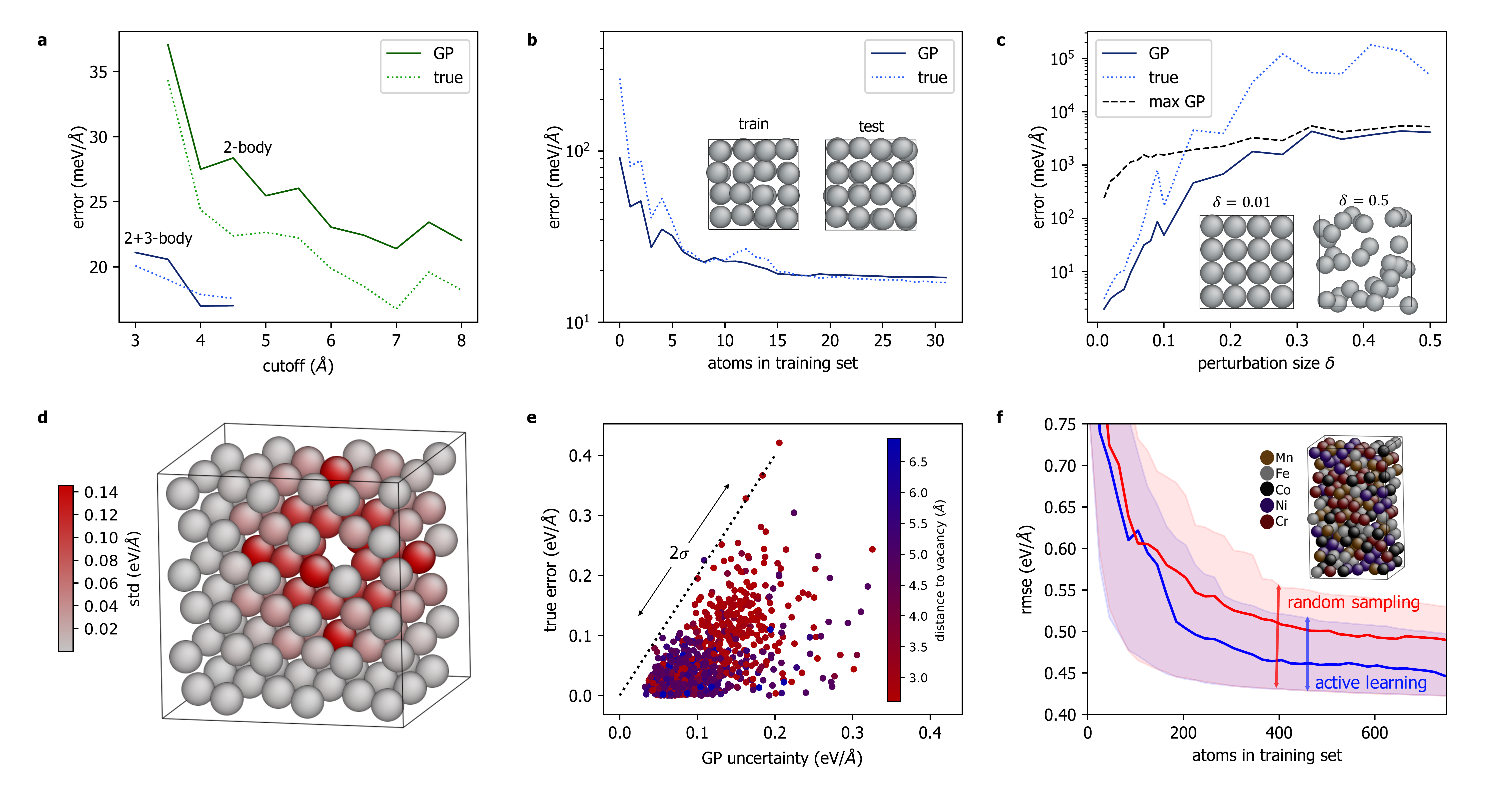}
	\caption{Tests of Gaussian process (GP) uncertainties. \textbf{a} Optimized noise uncertainty $\sigma_n$ (solid) and root mean squared error (RMSE, dotted) of GPs trained on an aluminum structure as a function of the cutoff radius $r_{\text{cut}}$ of the local environment for 2- and 2+3-body GP models (green and blue, respectively). \textbf{b} Mean GP uncertainty $\sqrt{\sigma_n^2 + \bar{\sigma}_{i\alpha}^2}$  (solid) versus true RMSE (dotted) as a function of the number of training environments. \textbf{c} Mean epistemic uncertainty $\bar{\sigma}_{i\alpha}$ (solid) and RMSE (dotted) on test structures with atomic coordinates perturbed from $\delta = 1\%$ to $50\%$ of the lattice parameter, with the upper bound on the epistemic uncertainty (dashed) approached for $\delta > 20\%$. \textbf{d} Uncertainties of individual force components for a GP model trained on bulk local environments. Each atom is colored by the most uncertain force component acting on the atom, with atoms closer to the vacancy having more uncertain forces. \textbf{e} Comparison of GP uncertainties and true model error for individual force components predicted on ten randomly perturbed Al vacancy structures, with most true errors falling within two standard deviations $\sigma = \sqrt{\sigma_n^2 + \sigma_{i\alpha}^2}$ of the predictive posterior distribution of the GP (dotted).
	\textbf{f} Learning curves of GP models trained on 5-element high entropy alloy structures, with training environments selected randomly (red) and with active learning (blue). The RMSE on force components of an independent test structure is plotted along with the distribution of uncertainties, shown as a band between the minimum and maximum uncertainties on force components in the structure.}
	\label{cal}
\end{figure*}

To justify an on-the-fly learning algorithm, we first characterize the noise and epistemic uncertainties of GP models constructed with the 2- and 3-body kernels described above, and compare them against test errors on out-of-sample structures. Importantly, the optimized noise uncertainty $\sigma_n$ and epistemic uncertainties $\sigma_{i\alpha}$ are found to provide a sensitive probe of true model error, with the noise uncertainty capturing the baseline error level of model predictions on local environments that are well represented in the training set, and the epistemic uncertainties capturing error due to deviation from the training data. In Fig.\ 2a-c, we test the relationship between GP uncertainties and true error by performing a set of plane-wave DFT calculations on a 32-atom supercell of FCC aluminum with the atoms randomly perturbed from their equilibrium sites. In Fig.\ \ref{cal}a, we examine the noise uncertainty $\sigma_n$ as a function of the cutoff radius of the model, which determines the degree of locality of the trained force field. 2- and 2+3-body GP models were trained on forces acting on atoms in a single structure and then tested on an independently generated structure, with the atomic coordinates in both cases randomly perturbed by up to $5\%$ of the lattice parameter, $a_{\text{lat}} = 4.046 \text{ \AA}$. For the 2-body models, the cutoff radius was swept from $3.5$ to $8$ \AA\ in increments of $0.5$ \AA, and for the 2+3-body models the 2-body cutoff was held fixed at $6$ \AA \ and the 3-body cutoff was swept from 3 to 4.5 \AA. The optimized noise uncertainty $\sigma_n$ plotted in Fig.\ \ref{cal}a closely tracks the root mean squared error (RMSE) on the test structure for the range of examined cutoff values.
The observed correlation provides a principled way to select the cutoff radius of the GP, showing that the expected error of a model with a given cutoff can be directly estimated from the optimized noise uncertainty $\sigma_n$ when the GP model has been trained on sufficient data.

When the GP model is trained on insufficient data, the epistemic uncertainties $\sigma_{i\alpha}$ rise above the noise uncertainty $\sigma_n$, indicating that the model requires additional training data to make accurate force estimates. The utility of the epistemic uncertainty is illustrated in Fig.\ \ref{cal}b, which examines GP uncertainties as a function of the amount of data in the training set. Using the same training and test structures as Fig.\ \ref{cal}a, a 2+3-body GP model with a $6 \text{ \AA}$ 2-body cutoff and $4 \text{ \AA}$ 3-body cutoff was constructed by adding local environments one by one to the training set and evaluating the RMSE and GP uncertainty after each update. The average GP uncertainty $\sqrt{\sigma_n^2 + \bar{\sigma}_{i\alpha}^2}$ closely tracks the RMSE, where $\bar{\sigma}_{i\alpha}$ is the mean epistemic uncertainty over all force components in the test structure.

We also demonstrate in Fig.\ \ref{cal}c that the epistemic uncertainty provides an accurate indicator of model error when the model is forced to extrapolate on local environments that are significantly different from local environments in the training set. To systematically investigate distance from the training set, a 2+3-body GP model was trained on a single aluminum structure with atomic coordinates perturbed by $\delta = 5\%$ of the lattice parameter and tested on structures generated with values of $\delta$ ranging from $1$ to $50\%$, with $\delta = 50\%$ giving rise to a highly distorted structure with a mean absolute force component of $28.6 \text{ eV/\AA}$ and a maximum absolute force component of $200.5 \text{ eV/\AA}$ (compared to a mean of $0.50 \text{ eV/\AA}$ and maximum of $1.48 \text{ eV/\AA}$ for the training structure). As shown in Fig.\ 2c, the mean epistemic uncertainty $\bar{\sigma}_{i\alpha}$ increases with $\delta$ and exceeds the optimized noise uncertainty of $\sigma_n = 11.53 \text{ meV/\AA}$ for $\delta > 5\%$, demonstrating the ability of the GP to detect when it is predicting on structures that are outside the training set. This capability is crucial for on-the-fly learning, as the model must be able to flag when additional training data is needed in order to accurately estimate forces. We furthermore observe that the error is substantially underestimated for large values of $\delta$ due to an upper bound on the epistemic uncertainty imposed by the signal variance hyperparameters of the kernel function, with the bound nearly saturated for $\delta > 20\%$ (see Methods for the definition of this bound). This emphasizes the importance of re-optimizing the hyperparameters when additional data is introduced to the training set, allowing the model to adapt to novel structures.

In Fig.\ \ref{cal}d and \ref{cal}e we demonstrate that GP uncertainties on individual force components can also provide valuable information about the expected errors on structures not represented in the training set. Fig.\ \ref{cal}d shows individual GP uncertainties $\sqrt{\sigma_{i\alpha}^2 + \sigma_n^2}$ on the predicted force components of a relaxed vacancy structure when the GP was trained on bulk local environments only. Each atom is colored according to the maximum uncertainty of the three predicted force components acting on the atom, with atoms closer to the defect tending to have higher uncertainties. This test was repeated for ten randomly perturbed vacancy structures, with the true error plotted in Fig.\ \ref{cal}e against the GP uncertainty $\sqrt{\sigma_{i\alpha}^2 + \sigma_n^2}$ of each force component, showing that higher uncertainties coincide with a wider spread in the true error.

We finally demonstrate in Fig.\ \ref{cal}f that the GP uncertainties are trustworthy for more complex multi-element systems. In this test, two GP models were trained on the five-element high entropy alloy (HEA) DFT forces of Ref.\ \cite{zhang2018deep}, with training environments selected randomly for the first GP model and with active learning for the second. Specifically, thirty-nine HEA structures were drawn from the ``rand 1'' portion of this dataset, and for each structure, twenty training environments were selected either at random or by identifying the highest uncertainty environments in the structure. After each update to the training set, both the GP uncertainties and true model error on an independent HEA structure were evaluated (with the test structure taken from the ``rand2'' portion of the dataset and having a different random allocation of elements). The distribution of total uncertainties $\sqrt{\sigma_{i\alpha}^2 + \sigma_n^2}$ on force components in the test structure is shown for both models in Fig.\ \ref{cal}f by plotting a band between the minimum and maximum uncertainties, which encloses the true RMSE. Actively selecting environments based on model uncertainty has the effect of shifting the learning curve downward, with the actively trained GP reaching a RMSE of $0.445 \text{ eV/\AA}$ on the test structure. The GP model obtained with active learning was subsequently mapped to a tabulated force field in order to rapidly evaluate forces on the entire ``rand2'' test set of Ref.\ \cite{zhang2018end}, which consisted of 149 HEA structures with elements placed at random lattice sites. The RMSE averaged over all test structures was found to be $0.466 \text{ eV/\AA}$ for the tabulated GP model, comparable to the RMSE of $0.410 \text{ eV/\AA}$ reported for the deep neural network model of Ref.\ \cite{zhang2018end} and outperforming the Deep Potential model of Ref.\ \cite{zhang2018deep}, which achieved a RMSE of $0.576 \text{ eV/\AA}$ on the same test set. We note that both neural network models were trained on $400$ HEA structures \cite{zhang2018end}, which exceeds the number of structures the GP was trained on by more than an order of magnitude.

\subsection{Aluminum crystal melt}

\begin{figure*}
	\centering
	\includegraphics[width=6.48in]{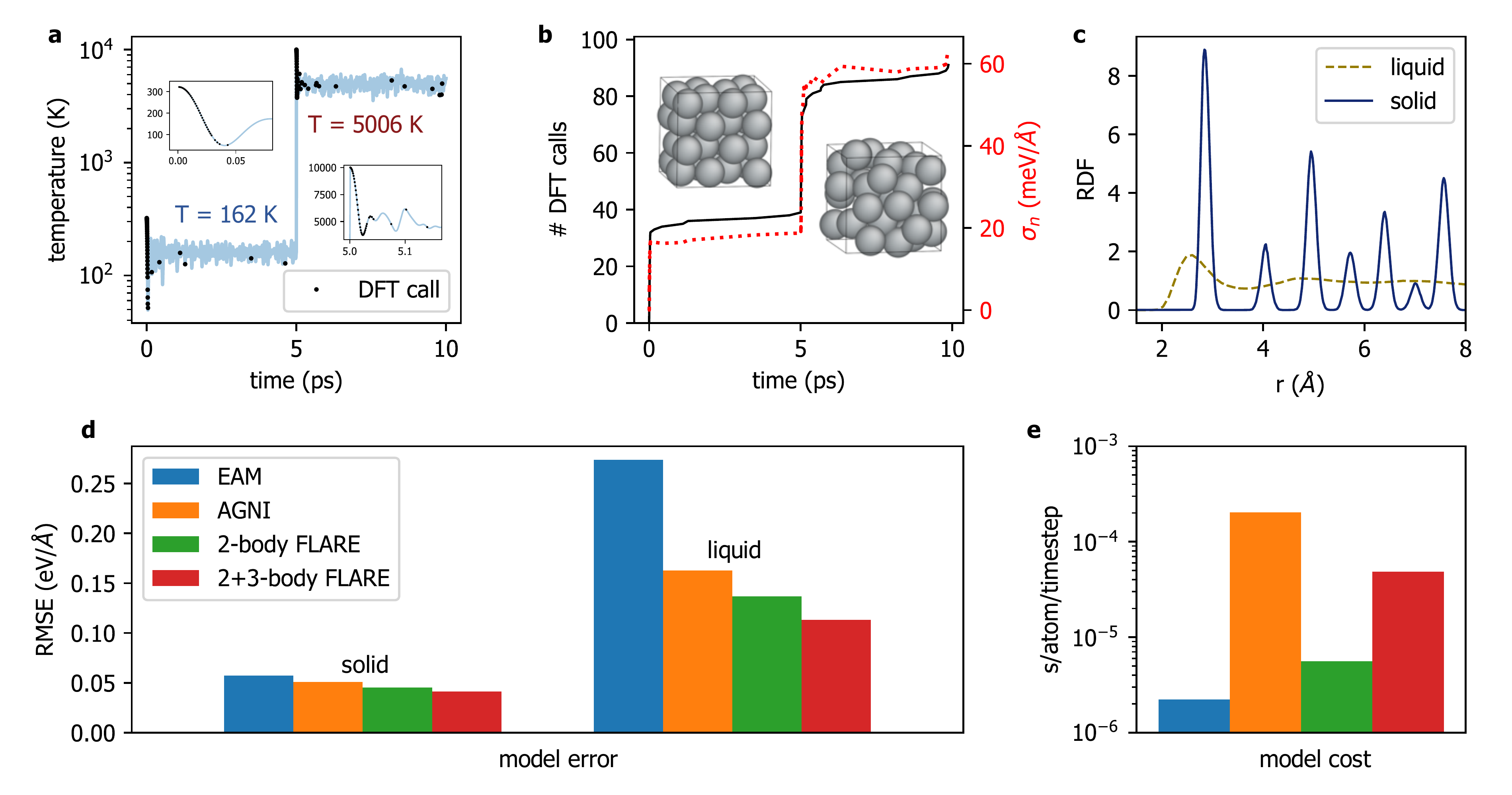}
	\caption{Active learning of a multi-phase aluminum force field. \textbf{a} Instantaneous temperature during a $10 \text{ ps}$ on-the-fly MD trajectory generated with the FLARE learning algorithm. The simulation begins in the FCC phase at low temperature and is melted at $t = 5 \text{ ps}$. When the epistemic uncertainty $\sigma_{i\alpha}$ on a force component rises above the current noise uncertainty $\sigma_n$ of the model, DFT is called (black dots). \textbf{b} The number of DFT calls (solid) and optimized noise uncertainty (dotted) throughout the simulation. A sharp increase is observed when the crystal is melted, illustrating the model's ability to actively learn the liquid phase. \textbf{c} During the first 5 ps of the simulation, the radial distribution function (RDF) is consistent with that of an fcc crystal (solid line), while in the final half of the simulation, the system exhibits an RDF characteristic of the liquid phase (dashed). \textbf{d} RMSE on AIMD forces of a tabulated version of the resulting force field compared with EAM, AGNI, and a tabulated 2-body FLARE force field. \textbf{e} Computational cost of LAMMPS implementations of these force fields on a single CPU core in s/atom/timestep.}
	\label{melt}
\end{figure*}

As a first demonstration of on-the-fly learning driven by GP uncertainties, we consider a 32-atom bulk aluminum system initialized in the FCC phase at low temperature, with $\mathcal{N_{\text{added}}} = 1$ local environment added to the training set whenever the epistemic uncertainty on a force component exceeds the current noise uncertainty, $\sigma_{\text{thresh}} = \sigma_n$. As shown in Fig.\ \ref{melt}a, DFT is called often at the beginning of the simulation as the GP model learns a force field suitable for FCC aluminum. After about 30 time steps, the model needs far fewer new training points, requiring fewer than 50 DFT calls in the first $5 \text{ ps}$ of the simulation. To test the model's ability to adapt to changing conditions, the crystal is melted at time $t = 5 \text{ ps}$ by rescaling the velocities of the atoms to give the system an instantaneous temperature of $10^4 \text{ K}$, well above the experimental melting point of aluminum (933 K) due to the strong finite size effects of the $2 \times 2 \times 2$ supercell. The subsequent temperature in the remaining $5$ ps of the simulation stabilizes around $5000$ K with a radial distribution function consistent with the liquid phase (Fig.\ \ref{melt}c). As shown in Fig.\ \ref{melt}b, which plots the cumulative number of DFT calls made during the training run, the GP model makes frequent calls to DFT immediately after the crystal melts, as the local environments in the liquid phase of aluminum are significantly different from the previous solid-state training environments. The noise uncertainty $\sigma_n$ of the model, shown in red in Fig.\ \ref{melt}b, sharply increases as the system enters the liquid phase, reflecting the fact that it is more difficult to model, involving more diverse local environments and significantly larger force fluctuations. Because the error threshold $\sigma_{\text{thresh}}$ is set equal to the optimized noise uncertainty $\sigma_n$, the threshold in the liquid phase is higher, and as a result the GP model requires a roughly similar number of DFT calls to learn the solid and liquid phases. Fewer than $100$ calls are needed in total during the 10 ps of dynamics, with the majority of DFT calls made at the beginning of the simulation and immediately after melting.

The obtained force field is validated by testing the model on two independent $10 \text{ ps}$ \textit{ab initio} molecular dynamics (AIMD) simulations of the solid and liquid phases of aluminum. $100$ structures were sampled from the AIMD trajectories with $0.1 \text{ ps}$ spacing between structures. Force predictions on all test structures were obtained with a tabulated version of the GP force field of Fig.\ \ref{melt}a and compared against the corresponding DFT values, with the RMSE in \text{eV/\AA} plotted in Fig.\ \ref{melt}d. For reference, the models are compared against state-of-the-art EAM and AGNI ML force fields, which were also trained on plane-wave DFT calculations with GGA exchange-correlation functionals and PAW pseudopotentials \cite{sheng2011highly, botu2016machine}, though we note that they were not trained on exactly the same DFT calculations as our models. Also included for comparison is the performance of a 2-body FLARE model trained on the same local environments as the 2+3-body model. Each force field was tested on the same structures, with the FLARE force field reaching the lowest force errors for both trajectories. This is due in part to the fact that FLARE optimizes the force field for the specific simulation of interest, only augmenting the training set when necessary. This bypasses the need to anticipate all possible phases which a system might explore when creating the force field. To assess computational efficiency, 1,000 MD steps were performed with the LAMMPS implementations of these four force fields on a single CPU core for a system of 1,372 bulk Al atoms, with the cost of each force field plotted in Fig.\ \ref{melt}e in s/atoms/timestep. The cost of the current LAMMPS implementation of the tabulated 2-body FLARE force field is found to be $5.6 \times 10^{-6} \text{ s/atom/timestep}$, which is the same order of magnitude as the EAM cost of $2.2 \times 10^{-6} \text{ s/atom/timestep}$. The 2+3-body model is about an order of magnitude slower at $4.9 \times 10^{-5} \text{ s/atom/timestep}$, but still faster than AGNI, which directly predicts forces with a small neural network. This makes FLARE considerably less expensive than many-body models like GAP, with the cost of the recent GAP silicon model reported as $0.1 \text{ s/atom/timestep}$ \cite{bartok2018machine}.


\begin{figure*}[h!]
	\centering
	\includegraphics[width=6.48in]{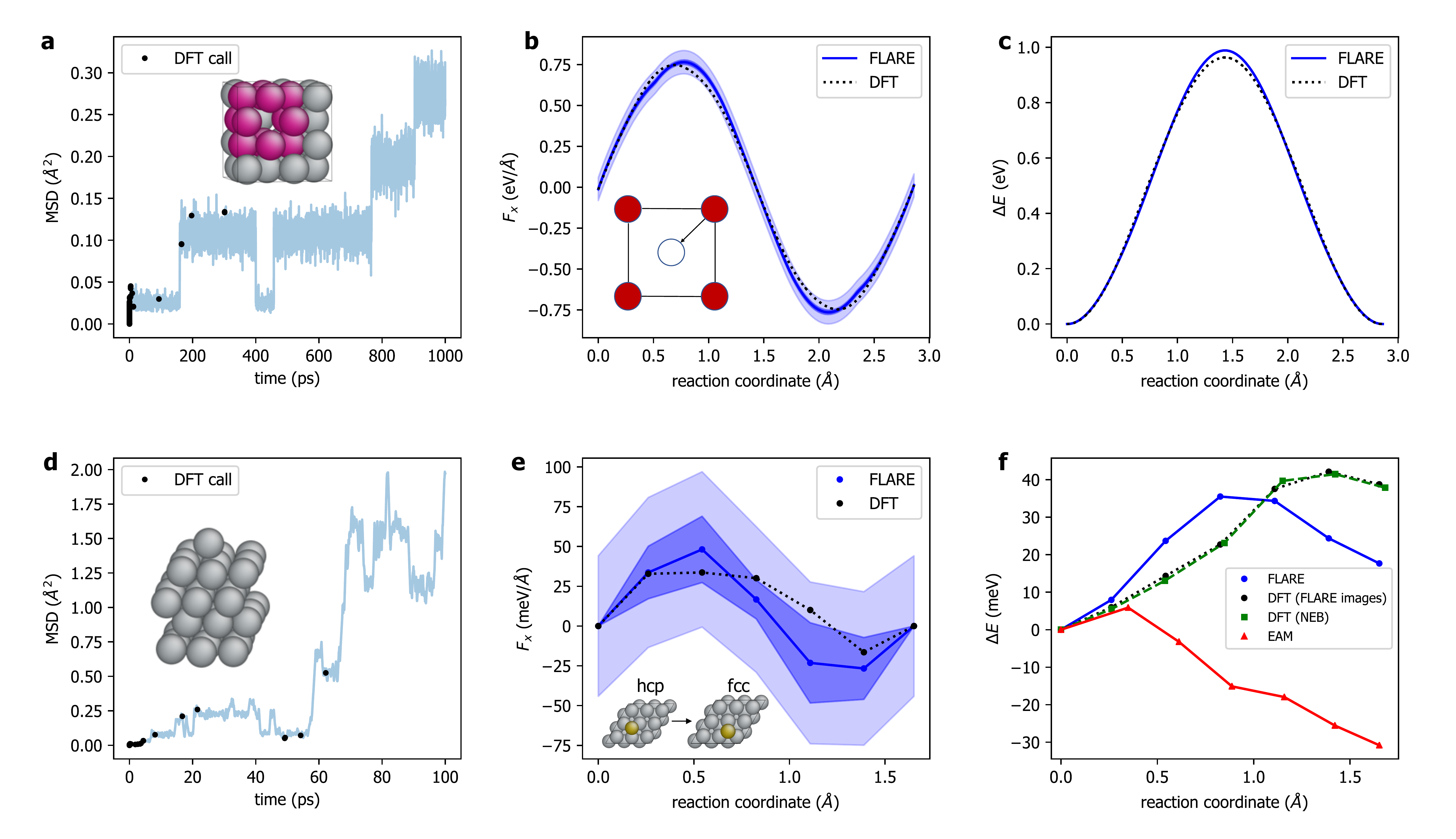}
	\caption{On-the-fly learning of vacancy and adatom diffusion in aluminum. \textbf{a} Mean squared displacement during a FLARE training run of duration 1 ns for a 31-atom fcc aluminum structure with a vacancy (see inset). The majority of DFT calls occur at the beginning of the run, with no additional calls required after the first 400 ps. \textbf{b} x-component of the force predicted by the resulting force field for vacancy migration along a high symmetry transition path (see inset), in close agreement with the ab initio barrier (dotted). \textbf{c} Predicted energies along the transition path (blue) compared with DFT (black). \textbf{d} On-the-fly learning of adatom diffusion on a (111) aluminum surface, with sharp jumps in the MSD signaling movement of the adatom on the surface. \textbf{e} x-component of the force on the adatom in nudged elastic band images of the hcp-to-fcc transition computed with the trained GP model (blue line), along with the epistemic uncertainty $\sigma_{ix}$ (dark blue) and total uncertainty $\sqrt{\sigma_{ix}^2 + \sigma_n^2}$ (light blue). DFT forces are computed for each image (black). \textbf{f} Predicted energies of the FLARE-generated NEB images (blue) relative to the first image. DFT energies are computed for each image (black), showing good agreement with the energies of an independent NEB calculation performed with DFT (green). NEB images with the EAM force field from \cite{sheng2011highly} are shown for comparison (red).}
	\label{diffusion}
\end{figure*}

\subsection{Bulk vacancy and surface adatom diffusion}
We next demonstrate that FLARE can be used to train force fields that dramatically accelerate simulations of rare-event dynamics over timescales spanning hundreds of picoseconds by applying the method to aluminum bulk vacancy diffusion and surface adatom diffusion. For bulk vacancy training, a 1 ns simulation was initialized by removing one atom from an equilibrium 32-atom FCC structure and setting the instantaneous initial temperature to $1500 \text{ K}$, giving a mean temperature of $734 \text{ K}$ across the simulation. The GP model was constructed with a 2-body kernel with cutoff $r_{\text{cut}}^{(2)} = 5.4 \text{ \AA}$, resulting in a final optimized noise uncertainty of $\sigma_n = 70.2 \text{ meV/\AA}$. Discarding the 3-body contribution was found to significantly accelerate the simulation while still achieving low force errors due to the simplicity of the single-defect bulk crystalline phase, opening up nanosecond timescales during training. As shown in Fig.\ \ref{diffusion}a, most DFT calls are made early on in the simulation, and after the first $\sim 400 \text{ ps}$, no additional DFT calls are required. The model predicts vacancy hops every few hundred picoseconds, which appear as sharp jumps in the mean squared displacement plotted in Fig.\ \ref{diffusion}a. To check the accuracy of the underlying energy model of the GP, DFT energies were computed along the high symmetry transition path sketched in the inset of Fig.\ \ref{diffusion}b, with a nearest neighbor migrating into the vacancy while all other atoms in the simulation cell were kept frozen at their fcc lattice sites. GP forces and energies along the transition path were evaluated to give an estimate of the energy barrier, showing close agreement with the \textit{ab initio} DFT values (Fig.\ \ref{diffusion}c), with the DFT forces lying within one standard deviation of the GP force predictions (Fig.\ \ref{diffusion}b). The entire FLARE training run, including DFT calculations, GP hyperparameter optimization, force evaluations and MD updates, were performed on a 32-core machine in 68.8 hours of wall time. Individual DFT calls required over a minute of wall time on average, making FLARE over 300 times faster than an equivalent AIMD run (see Supplementary Information for a breakdown of GP prediction costs).

To test the accuracy of FLARE on a subtler transition with a significantly lower energy barrier, we consider aluminum adatom diffusion on a four-layer (111) aluminum slab, with a representative structure shown in the inset of Fig.\ \ref{diffusion}d. As revealed in previous \textit{ab initio} studies, an isolated Al adatom on the (111) Al surface exhibits a small but surprising preference for the hcp site \cite{stumpf1994theory, stumpf1996ab}, making this system an interesting and challenging test for a machine learned force field. For this system, 3-body contributions were found to considerably increase the accuracy of the force field, with a $7 \text{ \AA}$ 2-body cutoff and $4.5 \text{ \AA}$ 3-body cutoff giving an optimized noise uncertainty of $\sigma_n = 44.2 \text{ meV/\AA}$ after the final DFT call at $t = 62.2 \text{ ps}$ (Fig.\ \ref{diffusion}d).
To validate the energetics of the force field, a 7-image nudged elastic band (NEB) calculation characterizing the transition from the hcp to fcc adatom sites was performed using the Atomic Simulation Environment \cite{larsen2017atomic} with the GP energy predictions shown in blue in Fig.\ \ref{diffusion}f. The DFT energies of each image of the NEB calculation are shown in black, showing agreement to within $\approx 20$ meV for each image and confirming the GP's prediction of a slight energetic preference for the hcp site in equilibrium, which is not reproduced by the EAM model of Ref.\ \cite{sheng2011highly} (red line in Fig.\ \ref{diffusion}f). An independent DFT NEB calculation was performed for the same transition, showing good agreement with the DFT energies of the FLARE NEB images.

\begin{figure*}[h!]
	\centering
	\includegraphics[width=6.48in]{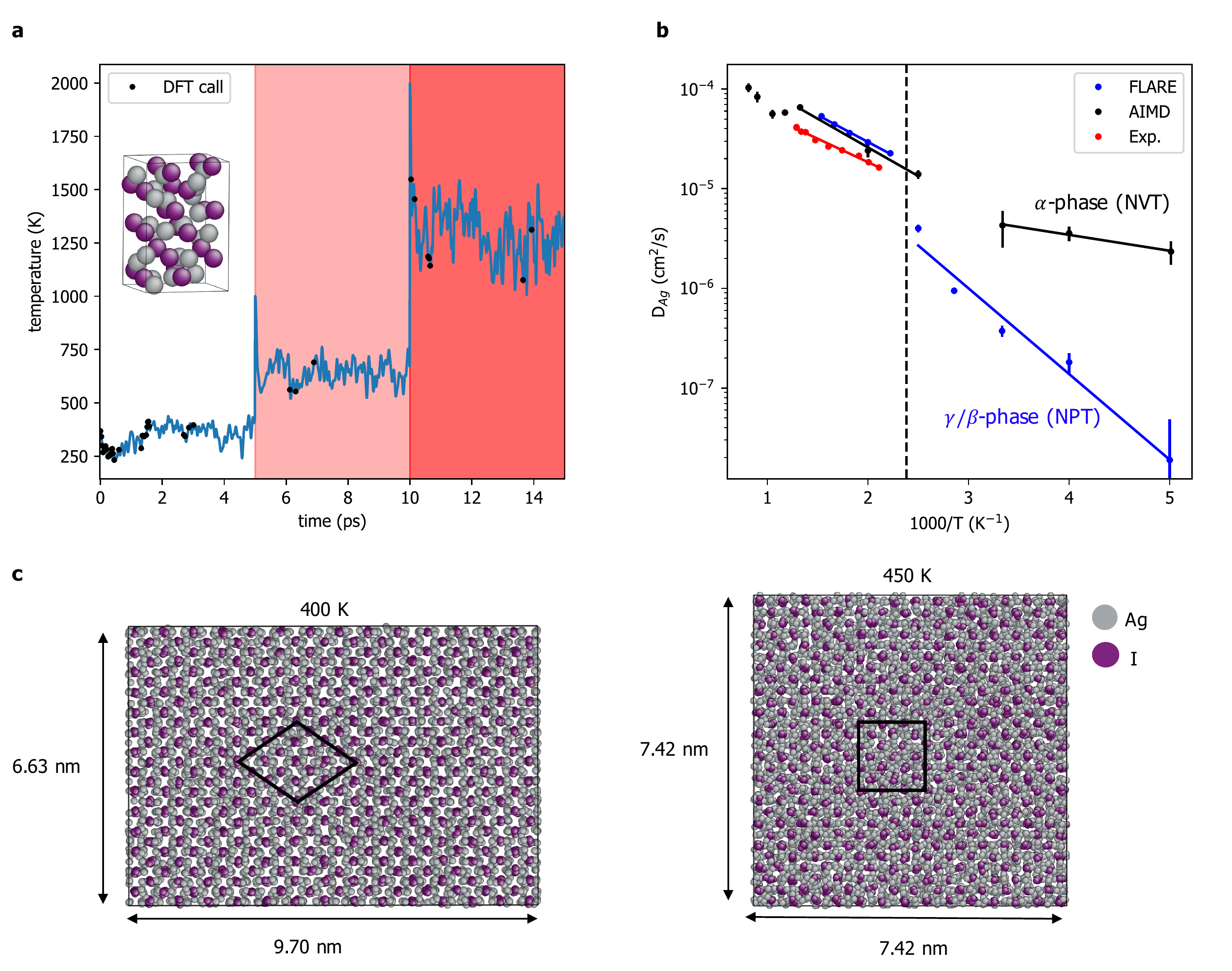}
	\caption{On-the-fly learning of fast-ion diffusion in silver iodide. \textbf{a} Temperature during a FLARE training simulation of duration 15 ps for a 48-atom silver iodide structure in the $\alpha$-phase (see inset), with the instantaneous temperature of the simulation increased at 5 ps and 10 ps. \textbf{b} Silver diffusion coefficients (blue dots) computed with a tabulated version of the resulting force field from $1 \text{ ns}$ NPT simulations of 10,976 AgI atoms. The computed coefficients for the $\alpha$-phase of AgI are in good agreement with the AIMD simulations of Ref. [1] (black) and experimental data reported in Ref. [2] (red), with the fast ion phase transition at 420 K (dashed line) correctly modeled. \textbf{c} Structures drawn from the 400 K (left) and 450 K (right) simulations, illustrating the solid-solid structural phase transition that occurs between these temperatures.}
	\label{iod}
\end{figure*}

\subsection{Fast-ion diffusion in AgI}
As a third and more challenging example of diffusion, we apply FLARE to the fast-ion conductor silver iodide (AgI), which exhibits a structural phase transition at $420 \text{ K}$ from the low-temperature $\gamma/\beta$-phase to a cubic ``superionic'' $\alpha$-phase, with silver ions in the $\alpha$-phase observed to have a liquid-like diffusivity \cite{hull2004superionics}. A 2+3-body FLARE model was trained in a 15 ps on-the-fly simulation of $48$ AgI atoms in the $\alpha$-phase, with the temperature increased at $5$ and $10 \text{ ps}$ (Fig.\ 5a). The uncertainty threshold was set to twice the noise uncertainty, $\sigma_{\text{thresh}} = 2 \sigma_n$, making the model slightly less sensitive to changing temperature and contributing to the $1 \text{ ps}$ delay observed between the temperature increase at $5 \text{ ps}$ and the next call to DFT at $t = 6.121 \text{ ps}$. Thirty-nine calls to DFT were made in total, with the $\mathcal{N}_{\text{added}}=10$ highest uncertainty local environments added to the training set after each DFT calculation.

After training, the model was mapped to a tabulated cubic spline model in LAMMPS, which was used to perform 1 ns simulations at zero pressure and fixed temperature, with each simulation requiring about three hours of wall time on 32 cpu cores ($\approx 3.2 \times 10^{-5} \text{ cpu $\cdot$ s/atom/timestep}$). Ten MD simulations were performed in total with temperatures ranging from $200$ to $650 \text{ K}$ in $50 \text{ K}$ intervals. In each simulation, the system was initialized in a pristine $14 \times 14 \times 14$ $\alpha$-phase supercell (10,976 atoms total), with the silver ions placed at the energetically preferred tetragonal interstices of the bcc iodine sublattice. The diffusion coefficients of the Ag ions are plotted in Fig.\ \ref{iod}b, showing a sharp increase between $400 \text{ K}$ and $450 \text{ K}$, in good agreement with the experimental fast-ion transition temperature of $420 \text{ K}$. The diffusion coefficients are compared with an AIMD study of the  $\alpha$-phase of AgI \cite{wood2006dynamical}, which used a similar exchange-correlation functional, showing excellent agreement at $450 \text{ K}$ and above. Both FLARE and AIMD show good agreement with experimentally observed $\alpha$-phase Ag diffusion coefficients \cite{kvist1970self}, with a slight vertical offset but comparable activation energies of activation energies of $0.107,$ $0.114,$ and $0.093 \text{ eV}$ for FLARE, AIMD, and experiment, respectively. Below the transition temperature, the FLARE force field correctly predicts a phase transition to a non-diffusive and non-cubic hcp phase with a nearest neighbor I-I coordination of $12$, consistent with the $\gamma$ and $\beta$ phases of AgI \cite{parrinello1983structural}. This accounts for the discrepancy between the FLARE and AIMD diffusion coefficients in the low temperature regime, as the latter simulations were conducted in the $\alpha$-phase with a fixed cubic cell. Example structures from the $400$ and $450 \text{ K}$ FLARE MD simulations are illustrated in Fig.\ \ref{iod}c, with the low temperature structure giving a $c/a$ ratio of $1.46$ and the high temperature structure having a lattice parameter of $a_{\text{lat}} = 5.30 \text{ \AA}$, in fair agreement with the corresponding experimental values of $c/a = 1.63$ and $a_{\text{lat}} = 5.07 \text{ \AA}$ near these temperatures (for the $\beta$- and $\alpha$-phases, respectively) \cite{collaboration1999silver}.

\subsection{General applicability} \label{app}
\begin{figure}
	\centering
	\includegraphics{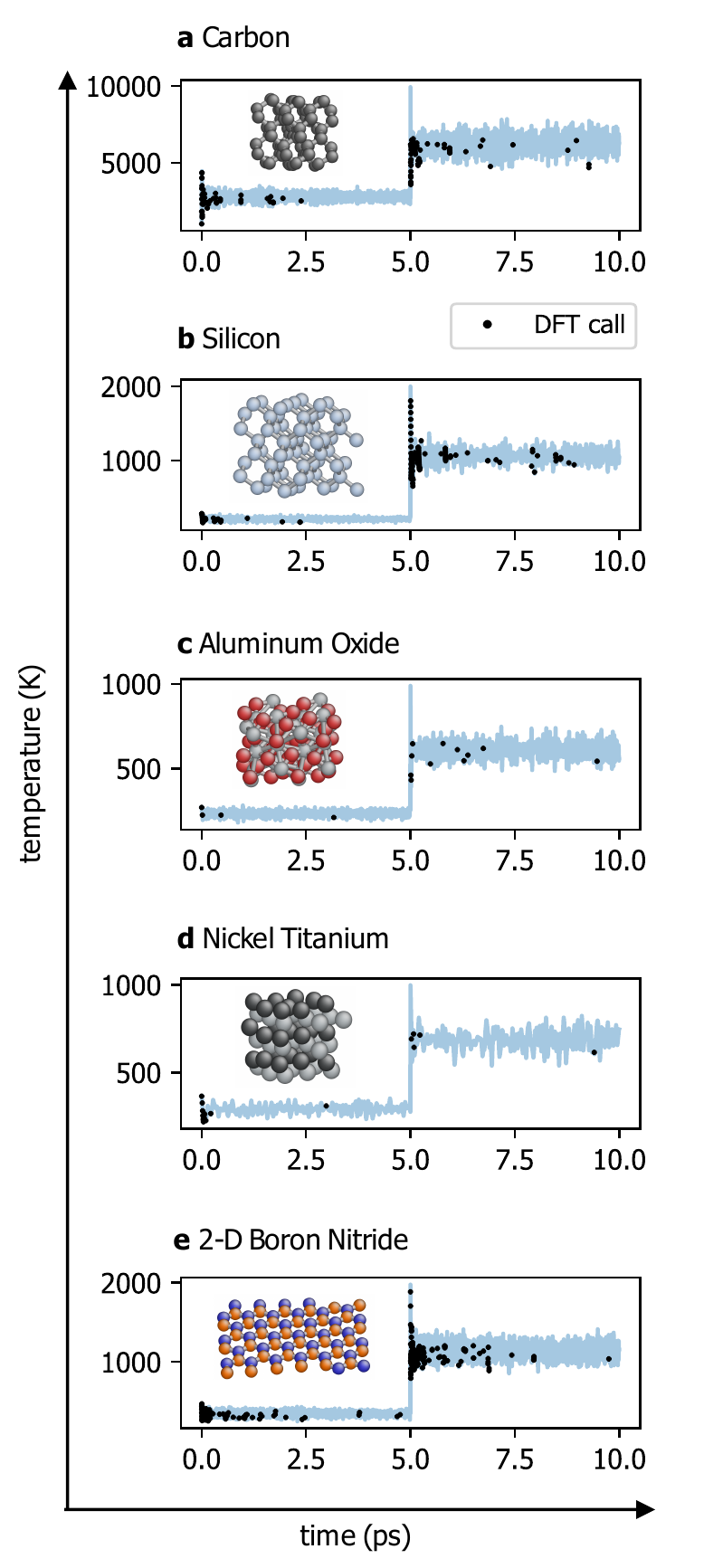}
	\caption{On-the-fly force field learning applied to a range of single- and multi-element systems. In each training run, the instantaneous temperature (blue) was increased at time $t = 5.0$ ps, triggering DFT calls and updates to the GP model (black dots) caused by model detection of novel local environments. Example structures from each simulation are shown in the insets.}
	\label{gen}
\end{figure}

Finally, we demonstrate in Fig.\ \ref{gen} that FLARE can be widely applied to diverse systems, including covalently bonded insulators and semiconductors, as well as oxides, alloys, and two-dimensional materials. FLARE training runs were performed for five representative systems---carbon, silicon, aluminum oxide, nickel titanium, and two-dimensional boron nitride---with the instantaneous temperature of each system rescaled at $t = 5$ ps to illustrate the model's ability to detect and adapt to novel local environments (see the left half of Table \ref{gen_tab} for training details). To accelerate training of the nickel titanium model, which required expensive DFT calculations, the error threshold was set to twice the noise uncertainty, $\sigma_{\text{thresh}} = 2 \sigma_n$, significantly reducing the total number of DFT calls needed  to $\sim 20$ (as shown in Fig.\ \ref{gen}d and Table \ref{gen_tab}). Adding multiple local environments to the training set after each DFT call also had the effect of reducing the total number of DFT calls needed, as apparent in the aluminum oxide training run, for which $\mathcal{N}_{\text{added}} = 30$ local environments were added after every DFT call and only $16$ DFT calls were needed in total to train the model. Each training run was performed on a 32-core machine and took between $11.3$ and $64.4$ hours of wall time (for silicon and carbon, respectively). We emphasize that the training procedure for each material is fully automatic, with the training set and hyperparameters updated on-the-fly without any human guidance.

To validate the models, independent NVE molecular dynamics trajectories of duration $10$ ps were generated with each GP force field, with DFT calculations performed for ten MD frames spaced equally across the simulation and compared against the corresponding GP predictions. We find low root mean squared errors (RMSE) of around $~0.1 \text{ eV/\AA}$ for four of the five systems, and for carbon we find a RMSE of $0.42 \text{ eV/\AA}$ due to the much higher temperature of the carbon validation run. The RMSE over all force component predictions in the ten representative frames is reported in Table \ref{gen_tab}. In order to illustrate the range of force magnitudes present in the simulation, we also report the $95$th percentile of the absolute force components in these frames, with the ratio of the two reported in the final column of Table \ref{gen_tab}. The resulting ratios lie between $3\%$ and $10\%$, similar to the ratios reported in a recent study of amorphous carbon with a Gaussian approximation potential \cite{deringer2017machine}.

\begin{table*}
    \centering
    \begin{tabular}{c | c c c c c c c |c c c c} 
    
    \hline
    \hline

    & \multicolumn{7}{c}{Training} & \multicolumn{4}{c}{Validation} \\
\hline

       & $\mathcal{N}_{\text{atoms}}$ & $r_{\text{2}}$ &  $r_{\text{3}}$ &  $\sigma_{\text{thresh}}$ & $\mathcal{N}_{\text{added}}$ & $\mathcal{N}_{\text{DFT}}$ & $t_{\text{wall}}$ & $T$ & RMSE & $P_{95}$ & Ratio \\ 

       &  & (\AA) & (\AA) &  &  &  & (hours) & (K) & (eV/\AA) & (eV/\AA) & \\ 
	 
	 \hline

	 C & $64$ & $4.0$ & $2.75$ & $\sigma_n$ & 1 & $107$ & $11.3$ & $3710$ & $0.42$ & $7.45$ & $0.056$ \\ 

     Si & $64$ & $6.0$ & $4.2$ & $\sigma_n$ & 5& $133$ & $64.4$ & $620$ & $0.077$ & $1.54$ & $0.050$ \\
     
     Al$_2$O$_3$ & $80$ & $4.5$  & $3.5$ & $\sigma_n$ & 30 & $16$ & $20.6$ & $533$ & $0.14$ & $1.85$ & $0.076$ \\

     NiTi & $54$ & $4.48$ & $3.2$ & $2 \sigma_n$ & 10 & $18$ & $33.3$ & $510$ & $0.10$ & $1.00$ & $0.104$ \\

     BN & $72$ & $5.1$ & $4.0$ & $\sigma_n$ & 1 & $237$ & $31.4$ & $677$ & $0.092$ & $3.24$ & $0.029$\\

	 \hline
	 \hline
	\end{tabular}
\caption{Training and validation details of the FLARE models shown in Fig.\ \ref{gen}. Training: $\mathcal{N}_{\text{atoms}}$ is the number of atoms in the training simulation, $r_2$ and $r_3$ are the 2- and 3-body cutoffs of the GP models, $\sigma_{\text{thresh}}$ is the uncertainty threshold that determines when DFT is called, $\mathcal{N}_{\text{added}}$ is the number of local environments added each time DFT is called, and $t_{\text{wall}}$ is the total wall time of the training simulation. Validation: $T$ is the mean temperature during the validation simulation, RMSE is the root mean squared error on ten snapshots from the validation run, and $P_{95}$ is the $95^{\text{th}}$ percentile of force components in these 10 snapshots. The ratio between the RMSE and $P_{95}$ is reported in the final column.}
\label{gen_tab}
\end{table*}

\section{Discussion}

In summary, we have presented a method for automatically training low-dimensional Gaussian process models that provide accurate force estimates and reliable internal estimates of model uncertainty. The model's uncertainties are shown to correlate well with true out-of-sample error, providing an interpretable, principled basis for active learning of a force field model during molecular dynamics. The nonparametric 2- and 3-body FLARE models described here require fewer training environments than high-dimensional machine learning approaches, and are therefore well-suited to settings where large databases of \textit{ab initio} data are too expensive to compute. Our models have a simple, accurate, and physically interpretable underlying energy model, which we have shown can be used to map the GP to a faster regression model approaching the speed of a classical force field. This provides a path toward force fields tailored to individual applications that give good agreement with DFT at several orders of magnitude lower computational cost, which we expect to considerably expand the range of materials that can be accurately studied with atomistic simulation. Particularly promising is the application of the FLARE framework to dynamical systems dominated by rare diffusion or reaction events, that are very difficult to treat with existing \textit{ab initio}, classical force field, or machine learning methods.

Extending this active learning method to complex systems like polymers and proteins is an important open challenge. The Bayesian force fields presented here may serve as a useful guide for selecting small, uncertain fragments from these large systems that can then be evaluated with DFT to refine the force field, similar to other recent approaches that train on small portions of large structures \cite{gastegger2017machine, mailoa2019fast}. This may provide a path toward accurate machine learned force fields for chemical and biological systems that are currently outside the reach of DFT and other quantum mechanical methods.

\section{Methods} \label{cov}

\subsection{Gaussian process force fields}
As observed by Glielmo \textit{et al.} \cite{glielmo2017accurate, glielmo2018efficient, glielmo2019building}, the task of fitting a force field can be dramatically simplified by assuming that only small clusters of atoms in the local environment of an atom $i$ contribute to its local energy $E_i$. We define the $n$-body local environment $\rho_i^{(n)}$ of atom $i$ to be the set of atoms within a cutoff distance $r_{\text{cut}}^{(n)}$ from atom $i$, and a cluster of $n$ atoms to be the atom $i$ and $n-1$ of the atoms in $\rho_i^{(n)}$.
The energy $\varepsilon_{\mathbf{s}_{i, i_1, ..., i_{n-1}}}(\mathbf{d}_{i, i_1, ..., i_{n-1}})$ of each cluster of $n$ atoms is assumed to depend on the species of the atoms in the cluster, $\mathbf{s}_{i, i_1, ..., i_{n-1}}=(s_i, s_{i_1}, .., s_{i_{n-1}})$, and on a corresponding vector of interatomic distances between the atoms, $\mathbf{d}_{i, i_1, ..., i_{n-1}}$. For example, for clusters of two atoms, this vector consists of a single scalar, $\mathbf{d}_{i, i_1} = (r_{i, i_1}),$ where $r_{i, i_1}$ is the distance between the central atom $i$ and atom $i_1$, and for clusters of three atoms, $\mathbf{d}_{i, i_1, i_2} = (r_{i, i_1}, r_{i, i_2}, r_{i_1, i_2})$. The local energy assigned to atom $i$ may then be written as
\begin{widetext}
\begin{equation}
E_i = \sum_{n=2}^{N} \sum_{i_{n-1} > ... > i_1 \in \rho_i^{(n)}} \varepsilon_{\mathbf{s}_{i, i_1, ..., i_{n-1}}}(\mathbf{d}_{i, i_1, ..., i_{n-1}}),
\label{loc_en}
\end{equation}
\end{widetext}
where the outer sum ranges over each $n$-body contribution to the energy up to a chosen maximum order $N$ and the inner sum ranges over all clusters of $n$ atoms inside the $n$-body environment $\rho_i^{(n)}$. The regression task is to learn the functions $\varepsilon_{\mathbf{s}_{i, i_1, ..., i_{n-1}}}(\mathbf{d}_{i, i_1, ..., i_{n-1}})$, which for small $n$ have much lower dimensionality than the full potential energy surface.

To learn the cluster contributions $\varepsilon_{\mathbf{s}_{i, i_1, ..., i_{n-1}}}$, we use \textit{ab initio} force data to construct Gaussian process (GP) models, an established Bayesian approach to describing probability distributions over unknown functions \cite{williams2006gaussian}. In GP regression, the covariance between two outputs of the unknown function is related to the degree of similarity of the inputs as quantified by a kernel function. For our GP force fields, the covariance between $n$-body energy contributions ($\varepsilon_{\mathbf{s}_{i, i_1, ..., i_{n-1}}}$ in Eq.\ (\ref{loc_en})) is equated to a kernel function $k_n$ that directly compares the interatomic distance vectors while preserving rotational invariance. The local energy kernel between two local environments $\rho_i, \rho_j$ is expressed as a sum over kernels between clusters of atoms,
\begin{widetext}
\begin{equation}
    k(\rho_i, \rho_j) = \sum_{n=2}^{N} \sum\limits_{\substack{i_{n-1} > ... > i_1 \in \rho_i^{(n)} \\ j_{n-1} > ... > j_1 \in \rho_j^{(n)}}} \sum_{\mathbf{P}_n} \delta_{\mathbf{s}_{i, i_1, ..., i_{n-1}}, \mathbf{P}_n \mathbf{s}_{j, j_1, ..., j_{n-1}}} k_n(\mathbf{d}_{i, i_1, ..., i_{n-1}}, \mathbf{P}_n \mathbf{d}_{j, j_1, ..., j_{n-1}}).
\label{kern}
\end{equation}
\end{widetext}
Importantly, this kernel function explicitly distinguishes between distinct species, with the delta function $\delta$ evaluating to $1$ if the species vectors $s_{i,i_1,...i_n}$ of the clusters under comparison are equal and $0$ otherwise. The innermost sum of Eq.\ (\ref{kern}) is over all permutations $\mathbf{P}_n$ of indices of the species and distance vectors of the second cluster, guaranteeing invariance of the model under permutation of atoms of the same species. The resulting force kernel describing the covariance between force components is obtained by differentiating the local energy kernel with respect to the Cartesian coordinates $\vec{r}_{i\alpha}, \vec{r}_{j\beta}$ of the central atoms of $\rho_1$ and $\rho_2$,
\begin{equation}
k_{\alpha,\beta}(\rho_i, \rho_j) = \frac{\partial^2 k(\rho_i, \rho_j)}{\partial \vec{r}_{i\alpha} \partial \vec{r}_{j\beta}},
\label{derv}
\end{equation}
giving an exactly rotationally covariant and energy conserving model of interatomic forces \cite{glielmo2017accurate, glielmo2018efficient, chmiela2017machine}. For completeness, we provide in Appendix B a table of formulas involved in computing the 3-body derivative kernel described by Eq.\ (\ref{derv}), along with its derivatives with respect to the hyperparameters of the kernel, which are used to calculate the gradient of the log marginal likelihood during hyperparameter optimization.

In this work, we choose $N = 3$, restricting the sum to 2- and 3-body contributions, as we have found the resulting GP models to be sufficiently expressive to describe with high accuracy a range of single- and multi-element systems while remaining computationally efficient. This is consistent with the findings of Ref.\ \cite{glielmo2018efficient}, which compared the performance of 2-, 3-, and many-body kernels and found that many-body models required substantially more training data while only modestly improving performance for several crystals, nanoclusters, and amorphous systems. Further investigation of model accuracy as a function of the maximum order $N$ of the kernel for different types of materials is an interesting area for future study, as it may provide a systematic data-driven approach to characterizing many-body interactions in complex materials.

For the pair and triplet kernels $k_2$ and $k_3$, we choose the squared exponential kernel multiplied by a smooth quadratic cutoff function $f_{\text{cut}}$ that ensures the model is continuous as atoms enter and exit the cutoff sphere,
\begin{widetext}
\begin{equation}
\begin{split}
k_2(r_{i, i_1}, r_{j, j_1}) &= \sigma_{s,2}^2 \exp\left(- \frac{(r_{i, i_1} - r_{j, j_1})^2}{2 \ell_{2}^2} \right) f_{\text{cut}}(r_{i, i_1}, r_{j, j_1}), \\
k_3(\mathbf{d}_{i, i_1, i_2}, \mathbf{d}_{j, j_1, j_2}) &= \sigma_{s, 3}^2 \exp\left(- \frac{||\mathbf{d}_{i, i_1,i_2} - \mathbf{d}_{j, j_1,j_2}||^2}{2 \ell_{3}^2} \right) f_{\text{cut}}(\mathbf{d}_{i, i_1,i_2}, \mathbf{d}_{j, j_1,j_2}),
\end{split}
\end{equation}
\end{widetext}
where $\sigma_{s, (2, 3)}$ is the signal variance related to the maximum uncertainty of points far from the training set, $\ell_{(2 ,3)}$ is the length scale of the 2- and 3-body contributions, and $||.||$ denotes the vector 2-norm.

The force component $f_{i\alpha}$ on each atom $i$ and the square of the epistemic uncertainty $\sigma_{i\alpha}^2$ assigned to that force component are computed using the standard GP relations \cite{williams2006gaussian},
\begin{equation}
	\begin{split}
f_{i\alpha} &= \bar{k}_{i\alpha}^{T} \left(K + \sigma_n^2 I \right)^{-1} \bar{y} \\
\sigma_{i\alpha}^2 &= k_{\alpha,\alpha}(\rho_i, \rho_i) - \bar{k}_{i\alpha}^{T} \left(K + \sigma_n^2 I \right)^{-1} \bar{k}_{i\alpha},
    \end{split}
\label{reg}
\end{equation}
where $\bar{k}_{i\alpha}$ is the vector of force kernels between $\rho_i$ and the local environments in the training set, i.e.\ $\bar{k}_{i\alpha, j\beta} = k_{\alpha, \beta}(\rho_i, \rho_j)$, $K$ is the covariance matrix $K_{m\alpha, n\beta} = k_{\alpha, \beta}(\rho_m, \rho_n)$ of the training points, $\bar{y}$ is the vector of forces acting on the atoms in the training set, and $\sigma_n$ is a hyperparameter that characterizes observation noise. The total uncertainty on the force component, corresponding to the variance of the predictive posterior distribution of the predicted value, is obtained by adding the square of the noise uncertainty $\sigma_n^2$ \cite{williams2006gaussian}. Notice that the square of the epistemic uncertainty is bounded above by $k_{\alpha, \alpha}(\rho_i, \rho_i)$, which for our kernel function is determined by the signal variances $\sigma_{s,2}^2$ and $\sigma_{s,3}^2$.

In all models in this work, the hyperparameters $\theta = \{\sigma_2, \sigma_3, \ell_2, \ell_3, \sigma_n \}$ are optimized with SciPy's implementation of the BFGS algorithm  \cite{scipy} by maximizing the log marginal likelihood of the training data $\rho = \{ \rho_1, \rho_2, ..., \rho_n\}$, which takes the form \cite{williams2006gaussian}
\begin{equation}
    \log p(\bar{y} | \rho, \theta) = -\frac{1}{2} \bar{y}^T (K + \sigma_n^2 I)^{-1} \bar{y} - \frac{1}{2} \log |K + \sigma_n^2 I| - \frac{n}{2} \log 2\pi.
\end{equation}
To efficiently maximize this quantity with BFGS, the gradient with respect to all hyperparameters is calculated with the analytic expression \cite{williams2006gaussian},
\begin{equation}
\frac{\partial}{\partial\theta_i} \log p(\bar{y}| \rho, \theta) = \frac{1}{2} \text{tr}\left( (\bar{\alpha} \bar{\alpha}^T - K^{-1}) \frac{\partial K}{\partial \theta_j} \right),
\end{equation}
where $\bar{\alpha} = K^{-1} \bar{y}$. The formulas for the kernel derivatives with respect to the hyperparameters that appear in this expression, $\frac{\partial K}{\partial \theta_j}$, can be exactly calculated, and we list them in Table III of the Supplementary Information for the case of the 3-body kernel. The BFGS algorithm is terminated once the log marginal likelihood gradient falls below a threshold value $\epsilon$ (in our implementation, we choose $\epsilon=10^{-4}$). Note that computation of the log marginal likelihood and its gradient involves inverting the covariance matrix $K$ and is efficient if the model is trained on fewer than $\sim 1000$ points. This data-driven approach to selecting model hyperparameters stands in contrast to other GP force fields, in which hyperparameters are chosen heuristically \cite{deringer2017machine}.

\subsection{Mapping to tabulated spline models}
As shown in Ref.\ \cite{glielmo2018efficient} for single-element systems, GP models built on $n$-body kernels can be mapped to efficient cubic spline models, eliminating the expensive loop over training points involved in the calculation of the kernel vector $\bar{k}_{i\alpha}$ in Eq.\ (\ref{reg}). We have extended this mapping procedure to our multi-element kernels by constructing cubic spline interpolants for each $n$-body force contribution $-\frac{d}{d\vec{r}_i}\varepsilon_{\mathbf{s}_{i, i_1, ..., i_{n-1}}}(\mathbf{d}_{i, i_1, ..., i_{n-1}})$. The $2$- and $3$-body contributions require $1$- and $3$-dimensional cubic splines, respectively. The resulting spline model can be made arbitrarily accurate relative to the original GP model by increasing the number of control points of the spline. In Table III of the Supplementary Information, we report the grid of control points used for each mapped force field in this work.

\subsection{Computational details}
All DFT calculations were performed using Quantum Espresso 6.2.1, with pseudopotentials, k-point meshes, plane-wave energy cutoffs, and charge density energy cutoffs for all calculations reported in Table I of the Supplementary Information. The on-the-fly learning algorithm is implemented with the FLARE package \cite{flare}, which couples our Python-based MD and GP code with Quantum ESPRESSO \cite{giannozzi2009quantum}. Kernel and distance calculations are accelerated with the open-source just-in-time compiler Numba to enable training simulations spanning hundreds of picoseconds \cite{lam2015numba}. All on-the-fly molecular dynamics trajectories were performed in the NVE ensemble using the Verlet algorithm. LAMMPS simulations of AgI were performed in the NPT ensemble at zero pressure. Atomistic visualizations were created using Atomeye \cite{li2003atomeye}.
\newpage
\section{Acknowledgements}
We thank Aldo Glielmo, Jin Soo Lim, Nicola Molinari, and Jonathan Mailoa for helpful discussions, and Anders Johansson, Kyle Bystrom, David Clark, and Blake Duschatko for contributions to the FLARE code. B.K. acknowledges generous gift funding support from Bosch Research and partial support from the National Science Foundation under Grant No. 1808162. L.S. was supported by the Integrated Mesoscale Architectures for Sustainable Catalysis (IMASC), an Energy Frontier Research Center funded by the U.S. Department of Energy, Office of Science, Basic Energy Sciences under Award \#DE-SC0012573. A.M.K. and S.B. acknowledge funding from the MIT-Skoltech Center for Electrochemical Energy Storage. S.B.T. is supported by the Department of Energy Computational Science Graduate Fellowship under grant DE-FG02-97ER25308.

\section{Competing interests}
\noindent
The authors declare no competing financial or non-financial interests.

\section{Data availability}
\noindent
The trained Gaussian process models and other simulation data are available from the authors upon request.

\section{Code availability}
\noindent
The FLARE code is available online at https://github.com/mir-group/flare.

\section{Author contributions}
\noindent
J.V.\ conceived the study and is the primary developer of the FLARE code. Y.X.\ implemented the mapping of the GP models to cubic splines, and L.S.\ created the LAMMPS pair style. Y.X., L.S., S.B.T.\ and S.B.\ assisted with code development. B.K.\ supervised the work and contributed to algorithm development. J.V. wrote the manuscript. All authors contributed to manuscript preparation.

\bibliography{bib.bib}

\end{document}


\title{On-the-Fly Active Learning of Interpretable Bayesian Force Fields for Atomistic Rare Events: Supplementary Information}

\author{Jonathan Vandermause}
\author{Steven B. Torrisi}
\author{Simon Batzner}
\author{Yu Xie}
\author{Lixin Sun}
\author{Alexie M. Kolpak}
\author{Boris Kozinsky}

\maketitle

\begin{table*}[h!]
	\centering
	\begin{tabular}{c c c c c} 
	 \hline
	 \hline
	  Material & Pseudopotential & K-points &  Energy (Ry) & Density (Ry) \\ 
	 \hline 
	 Al (Fig.\ 2) & Al.pbe-n-kjpaw-psl.1.0.0.UPF & 7x7x7 & 29 & 143 \\ 
     Al (Figs. 4-5) & Al.pbe-n-kjpaw-psl.1.0.0.UPF & 2x2x2 & 29 & 143 \\
     Al (Fig. 6) & Al.pbe-n-kjpaw-psl.1.0.0.UPF & 4x4x2 & 29 & 143 \\
     AgI & \Bigg\{\makecell{Ag.pbe-n-kjpaw\_psl.1.0.0.UPF \\ I.pbe-n-kjpaw\_psl.1.0.0.UPF} & $\Gamma$ & 45 & 181 \\
     C & C.pbe-n-kjpaw\_psl.1.0.0.UPF & 2x2x2 & $40$ & $326$ \\
     Si & Si.pbe-n-kjpaw\_psl.1.0.0.UPF & $\Gamma$ & 44 & 175 \\
     Al$_2$O$_3$ & \Bigg\{\makecell{Al.pbe-n-kjpaw-psl.1.0.0.UPF \\ O.pbe-n-kjpaw\_psl.1.0.0.UPF} & $\Gamma$ & 47 & 323 \\
     NiTi& \Bigg\{\makecell{Ni.pz-n-rrkjus\_psl.0.1.UPF \\ Ti.pz-spn-rrkjus\_psl.1.0.0.UPF} & 2x2x2 & 52 & 576 \\
     BN & \Bigg\{ \makecell{B.pbe-n-kjpaw\_psl.1.0.0.UPF \\ N.pbe-n-kjpaw\_psl.1.0.0.UPF} & $\Gamma$ & 44 & 325 \\
	 \hline
	 \hline
	\end{tabular}
\caption{Details of the DFT calculations performed in this work. All pseudopotentials are available in the Quantum Espresso pseudopotential library \cite{dal2014pseudopotentials, pseudo}.}
\label{det}
\end{table*}

\section{Kernel evaluation cost}
To provide an estimate of the computational cost of evaluating forces with our Python-based GP models before mapping them to cubic splines, we provide empirical wall times of 2- and 2+3-body kernel evaluations and GP force and variance predictions in Table \ref{time} below. The test was performed with the 2- and 2+3-body aluminum melt GP models presented in Fig.\ 3 of the main text. All tests were performed on the same atomic environment $\rho_{\text{test}}$ taken from the training set of the GP, which contained $42$ atoms inside the $2$-body cutoff sphere and $18$ atoms inside the $3$-body cutoff sphere. We report in the first two rows the wall time of evaluating $k_{y, z}(\rho_{\text{test}}, \rho_{\text{test}})$ for the 2- and 2+3-body kernels on a single CPU core, showing that the 2+3-body kernel requires about an order of magnitude more wall time due to the double loop over neighbors that is required. The next two rows report the cost of predicting the $x$-component of the force on $\rho_{\text{test}}$ along with the epistemic uncertainty $\sigma_{ix}$ on this force component. The time is approximately a factor of $93 \times 3 = 279$ larger than the corresponding kernel evaluation times, due to the loop over training labels required in the evaluation of the GP mean and variance (Eq.\ (5) of the main text). We also report in the final two rows the cost of evaluating the mean prediction alone, which is only slightly less than the cost of evaluating both the mean and variance. This is because the majority of the computation time in both cases is spent evaluating the kernel vector $\bar{k}_{i\alpha}$ in Eq.\ (5), not in performing the matrix/vector multiplications.

\begin{table}[h!]
    \centering
    \begin{tabular}{| c || c |}
        \hline
        Computed Quantity & Time (ms)\\
        \hline
        \hline
        2-body kernel & 0.039 \\
        \hline
        2+3-body kernel & 0.46 \\
        \hline
        \hline
        2-body prediction (M+V) & 11.36 \\
        2+3-body prediction (M+V) & 138.72 \\
        \hline
        \hline
        2-body prediction (M) & 10.90 \\
        2+3-body prediction (M) & 136.71 \\
        \hline
    \end{tabular}
    \caption{Wall times of kernel evaluations and GP force and variance predictions. Reported wall times were averaged over $1000$ evaluations.}
    \label{time}
\end{table}

\section{Cubic spline models}
In the table below, we report the grid of control points used to construct the spline-based tabulated force fields discussed in the main text.

\begin{table}[h!]
    \centering
    \begin{tabular}{| c | c | c|}
        \hline
        Model & $2$-body & $3$-body \\
        \hline
        \hline
        High entropy alloy (Fig.\ 2f) & $100$ & $30 \times 30 \times 30$ \\
        \hline 
        $2$-body Aluminum (Fig.\ 3) & $100$ & - \\
        \hline
        $2+3$-body Aluminum (Fig.\ 3) & $100$ & $15 \times 15 \times 15$ \\
        \hline
        Silver iodide  (Fig.\ 5) & 64 & $15 \times 15 \times 15$ \\
        \hline

    \end{tabular}
    \caption{Cubic spline grids of the mapped force fields discussed in the main text.}
    \label{grids}
\end{table}

\section{Derivative kernel expressions}
Expressions used to compute the three-body force kernel defined in Eq.\ (3) of the main text are reported in Table \ref{kern_exp}.
\begin{center}
    \begin{table*}
    \begin{tabular}{ |c|c|c| } 
     \hline
     Energy Kernel & $k_3(\mathbf{d}_{i, i_1, i_2}, \mathbf{d}_{j, j_1, j_2})$ & $\sigma_{s,3}^2 k_{\text{SE}}(\mathbf{d}_{i, i_1, i_2}, \mathbf{d}_{j, j_1, j_2}) f_{\text{cut}}(\mathbf{d}_{i, i_1, i_2}) f_{\text{cut}}(\mathbf{d}_{j, j_1, j_2})$ \\ 
     \hline
     - & $k_{\text{SE}}(\mathbf{d}_{i, i_1, i_2}, \mathbf{d}_{j, j_1, j_2})$ & $\exp\left( - \frac{||\mathbf{d}_{i, i_1, i_2} - \mathbf{d}_{j, j_1, j_2} ||^2}{2 \ell^2} \right)$\\
     \hline
     - & $\mathbf{d}_{j, j_1, j_2}$ & $(r_{i, i_1}, r_{i, i_2}, r_{i_1, i_2})$ \\
     \hline
     - & $f_{\text{cut}}(\mathbf{d}_{i, i_1, i_2})$ & $f(r_{i, i_1}) f(r_{i, i_2}) f(r_{i_1, i_2})$ \\
     \hline
      - & $f(r)$ & $(r - r_{\text{cut}})^2$ \\
     \hline
    Force Kernel & $\frac{\partial^2 k_3}{\partial \vec{r}_{i \alpha} \partial  \vec{r}_{j \beta}}$ & $\sigma^2_{s,3} (k_0 + k_1 + k_2 + k_3)$ \\ 
     \hline
     - & $k_0$ & $k_{\text{SE}} \frac{\partial f_{\text{cut}}(\mathbf{d}_{i, i_1, i_2})}{\partial  \vec{r}_{i \alpha}} \frac{\partial f_{\text{cut}}(\mathbf{d}_{j, j_1, j_2})}{\partial  \vec{r}_{j \beta}}$ \\ 
     \hline
     - & $k_1$ & $\frac{\partial k_{\text{SE}}}{\partial  \vec{r}_{i \alpha}} f_{\text{cut}}(\mathbf{d}_{i, i_1, i_2}) \frac{\partial f_{\text{cut}}(\mathbf{d}_{j, j_1, j_2})}{\partial  \vec{r}_{j \beta}}$ \\ 
     \hline
     - & $k_2$ & $\frac{\partial k_{\text{SE}}}{\partial  \vec{r}_{j \beta}} \frac{\partial f_{\text{cut}}(\mathbf{d}_{i, i_1, i_2})}{\partial  \vec{r}_{i \alpha}} f_{\text{cut}}(\mathbf{d}_{j, j_1, j_2})$ \\ 
     \hline
     - & $k_3$ & $\frac{\partial^2 k_{\text{SE}}}{\partial  \vec{r}_{i \alpha} \partial  \vec{r}_{j \beta}} f_{\text{cut}}(\mathbf{d}_{i, i_1, i_2}) f_{\text{cut}}(\mathbf{d}_{j, j_1, j_2})$ \\ 
     \hline
     - & $\frac{\partial k_{\text{SE}}}{\partial  \vec{r}_{i \alpha}}$ & $\frac{k_{\text{SE}} B_1}{\ell^2}$ \\ 
     \hline
     & $B_1$ & $\frac{(r_{i, i_1} - r_{j, j_1}) (\vec{r}_{i_1 \alpha} - \vec{r}_{i \alpha})}{r_{i, i_1}} + \frac{(r_{i, i_2} - r_{j, j_2}) (\vec{r}_{i_2 \alpha} - \vec{r}_{i \alpha})}{r_{i, i_2}}$ \\
     \hline
     - & $\frac{\partial k_{\text{SE}}}{\partial  \vec{r}_{j \beta}}$ &  $-\frac{k_{\text{SE}} B_2}{\ell^2}$ \\ 
     \hline
    - & $B_2$ & $\frac{(r_{i, i_1} - r_{j, j_1}) (\vec{r}_{j_1 \beta} - \vec{r}_{j \beta})}{r_{j, j_1}} + \frac{(r_{i, i_2} - r_{j, j_2}) (\vec{r}_{j_2 \beta} - \vec{r}_{j \beta})}{r_{j, j_2}}$ \\
     \hline
     - & $\frac{\partial^2 k_{\text{SE}}}{\partial  \vec{r}_{i \alpha}  \vec{r}_{j \beta}}$ & $\frac{k_{\text{SE}}}{\ell^4} \left(A \ell^2 -B_1 B_2 \right)$\\ 
     \hline
     - & $A$ & $ \frac{ (\vec{r}_{i_1 \alpha} - \vec{r}_{i \alpha})  (\vec{r}_{j_1 \beta} - \vec{r}_{j \beta})}{r_{i, i_1} r_{j, j_1}} + \frac{ (\vec{r}_{i_2 \alpha} - \vec{r}_{i \alpha})  (\vec{r}_{j_2 \beta} - \vec{r}_{j \beta})}{r_{i, i_2} r_{j, j_2}}$ \\
     \hline
     $\ell$ Derivative & $\frac{\partial^3 k_{3}}{\partial \ell \partial  \vec{r}_{i \alpha} \partial  \vec{r}_{j \beta}}$ & $\sigma_{s,3}^2  \left(\frac{\partial k_0}{\partial \ell} + \frac{\partial k_1}{\partial \ell} + \frac{\partial k_2}{\partial \ell} + \frac{\partial k_3}{\partial \ell}\right)$ \\
     \hline
     - & $\frac{\partial k_{\text{SE}}}{\partial \ell}$ & $\frac{k_{\text{SE}} ||\mathbf{d}_{i, i_1, i_2} - \mathbf{d}_{j, j_1, j_2}||^2}{l^3}$ \\
     \hline
     - & $\frac{\partial^2 k_{\text{SE}}}{\partial \ell \partial  \vec{r}_{i \alpha}}$ & $B_1 \left( \frac{1}{\ell^2} \frac{\partial k_{\text{SE}}}{\partial \ell} - \frac{2 k_{\text{SE}}}{\ell^3} \right)$ \\
     \hline
     - & $\frac{\partial^2 k_{\text{SE}}}{\partial \ell \partial  \vec{r}_{j \beta}}$ & $-B_2 \left( \frac{1}{\ell^2} \frac{\partial k_{\text{SE}}}{\partial \ell} - \frac{2 k_{\text{SE}}}{\ell^3} \right)$ \\
     \hline
     - & $\frac{\partial^3 k_{\text{SE}}}{\partial \ell \partial  \vec{r}_{i \alpha} \partial  \vec{r}_{j \beta}}$ & $\left( A \ell^2 - B_1 B_2 \right) \left( \frac{\partial k_{\text{SE}}}{\partial \ell} \frac{1}{\ell^4} - \frac{4 k_{\text{SE}}}{\ell^5} \right) + \frac{2 k_{\text{SE}} A}{\ell^3}$ \\
     \hline
     $\sigma$ Derivative & $\frac{\partial^3 k_3}{\partial \sigma_{s,3} \partial  \vec{r}_{i \alpha} \partial  \vec{r}_{j \beta}}$ & $2 \sigma_{s, 3} (k_0 + k_1 + k_2 + k_3)$ \\
     \hline
    \end{tabular}
    \caption{Quantities used to calculate the three-body force kernel and its derivatives with respect to the hyperparameters $\ell$ and $\sigma$, which are required in the calculation of the log marginal likelihood. Terms in the second column are defined in the third column.}
    \label{kern_exp}
\end{table*}
\end{center}

\bibliography{bib.bib}